\documentclass[a4paper,12pt]{article}

\usepackage{epsfig}

\begin{document}

\title{Evidence of Non-Zero Mass Features for the Neutrinos  
Emitted at Supernova LMC-'87A}
\author{ Humiaki  Huzita\\
 INFN, Sez. di Padova, I-35131,  Padova,  Italy 
\\
	                        e-mail     huzita@padova.infn.it
} 
 \date{\today}
  \maketitle

\begin{abstract}
The observation of the neutrinos arrived from Supernova LMC-'87A shows, with a good   
confidence level, the existence of two massive neutrinos. For the unobserved third neutrino 
mass,  could speculate two possibilities either that this mass is close to one of the two observed values or 
that this neutrino has a negligible electronic flavor component.  
 \end{abstract}

\section{Introduction} 
	The explosion of a faraway supernova is an event quite suitable for understanding 
some of the most important physical features of neutrinos.  After being produced, neutrinos 
pass through dense matter and therefore both their initial flavors and masses might 
considerably change. However, in their subsequent long journey through a good vacuum 
towards the Earth, no further interaction will practically occur, and their mass states will not 
change. If the masses of the three kinds of neutrinos are not zero and differ from each other 
more than order of eV, the wave packet of each mass state will separate from the others much 
more than the Earth dimension and there would be almost no interference between states of 
different masses. The detector can see all the three flavors for each separate mass group 
provided the neutrino energy is high enough. Therefore a complete knowledge of the basic 
features of neutrinos, {\it i.e.} the mass values and the mixing angles to all the flavor states from 
each separated mass state, could be obtained without too big uncertainty.  

	The explosion of Supernova LMC-'87A has been a lucky event, since it occurred at 
a distance of $\sim$ 200,000 light-years and the produced neutrinos arrived  on 
the Earth in 1987, when three huge detectors, able to observe neutrino
interactions, were 
already operative. Moreover, the distance is large enough to separate the neutrino mass states 
(provided these differ more than order of eV) and, at the same time, is not so large to make the 
statistics of the observed events too poor for a quantitative analysis.  

	The observations made at Kamiokande II \cite{1,2}, IMB
\cite{3,4} and Baksam \cite{5} show an 
apparent disagreement with the theory:  the number of the observed events is roughly one order of 
magnitude smaller, while the spread in the arrival times is about one order of magnitude 
larger than the values, both comparing to the values expected on a theoretical ground \cite{6}. 
Thus, the first disagreement
indicates an explosion smaller than the predicted one, while the second
indicates a bigger. This contradiction disappears if neutrinos are massive particles. In this case, a massive 
neutrino with a lower energy travels more slowly than a higher energy one and will arrive 
on the Earth appreciably later than the latter due to long distance the neutrinos must go 
through.  In this way, the large spread observed in the arrival times could be easily 
understood. A further consequence of the assumption that neutrinos are massive particles is 
the fact that the plot of the arrival time of each neutrino {\it versus} the reciprocal of the square of 
the observed energy must show a grouping of the points along three different straight lines, 
whose slopes are proportional to the squared mass of each kind of neutrino (see sect. 5). The 
plot is given in Fig.3. It shows only two linear groupings indicating the mass values $3.4 \pm 0.6$ 
and $22.7 \pm 3.7$ eV.  One could speculate on the reasons hiding the third mass value (or the 
third linear grouping). This mass could be close to one of the two observed mass values or the 
electronic flavor of the third neutrino could be so small to yield visible events. In the latter 
case its mass value is completely unknown and any positive value, including zero, is possible. 
 
\section{Discrepancy between expectation and observation by Kamiokande II }
Due to the fact that the neutrinos arriving from Supernova LMC-'87A do not have a 
very high energy, their observation was made possible by detecting the electrons produced  
by the neutrino charged current interactions in the detector. The observed electron energy 
gives a good estimate of the primary neutrino energy while the neutrinos' arrival times are 
measured with a precision higher than a millisecond. Therefore we have two well observed 
quantities for each event: the neutrino energy and the arrival time. 

	According to theoretical predictions, the Kamiokande II detector in Japan (3 kilo 
tons of water) should have observed, within a few seconds or less, nearly 50 events with 
energy higher than 10 MeV, with an average energy of $10 \sim 15$ MeV. The real events observed 
at Kamiokande II were in total 12 with an average energy of 14.6 MeV, and only 7 of these 
had an energy higher than 10 MeV. Further, the total time spread between the first and the 
last observed events was found to be equal to 12.44 seconds.  

	Considering the lowest energy value (6.3 MeV) of the observed events as the 
detector's threshold energy and the fact that around this energy the detector efficiency is still 
rather low, the real average energy should be lower than 14.6 MeV. It could
likely be about
10 MeV, or even lower. The observed small number of events with energy greater than 10 
MeV, where the detector's efficiency should already be fairly good, indicates that the real 
supernova explosion was energetically less powerful than theoretically expected. On the 
contrary, the wide time spread of the observed events appears to indicate that the real 
explosion was much bigger, so as to have a longer emission of neutrinos. In fact, according 
to estimations, the majority of events were expected to occur within one second and only a 
negligible fraction was expected after a few of seconds (see Fig.1) 
\cite{9}. 

	The two discrepancies between the observed and predicted number of events and the 
observed and predicted spread of their arrival times appear to be contradictory. But, as 
already anticipated, this inconsistency is only apparent since it reflects the fact that neutrinos 
are massive particles, as it will appear clear in the following. In fact, while massless neutrinos 
propagate with the light speed and their time distribution does not change whatever the 
distance they go through, for massive neutrinos the time spread in the arrival times does no 
longer depend only on the spread in their production times but also on the time delay for 
traveling the distance Supernova-Earth. Thus, for each neutrino, this arrival 
time depends both on 
its mass and its energy, since the velocity of a massive particle depends on these two 
quantities.    
 
\section{Comparison of the Kamiokande data with that of the IMB and the Baksam}  
	The Kamiokande II data can be compared with those observed by the IMB detector 
in USA and the Baksan detector in Caucasus, which have different threshold energies, 
because the time distribution of events does neither depend on the detectors' working 
mechanism nor on their dimensions, responsible only of  their statistical efficiency. The 
correlation between the spread time ({\it i.e.} the time difference between the first and the last 
arrived events) and the lowest energy of the observed events (related to the detector's 
threshold) is shown in the following table and in Fig.2.  
 \begin{center}
  \begin{tabular}{llclc}
   &&\multicolumn{1}{r}{The lowest energy(MeV)}&&\multicolumn{1}{r}{ Spread time
(s)} \\
        Kamiokande II     &&                  6.3          &&      12.4\\
     Baksan          &&                12.0           &&        9.1\\
     IMB             &&                20.0           &&       5.6\\
   \end{tabular}
  \end{center}

In Fig.2, for each detector the error bars were obtained using the differences between the 
observed lowest energy values and between the observed latest arrival times, respectively. 
The figure shows that the observed time spreads depend almost linearly on the threshold 
energies of the detectors.

	Even if the neutrinos with lower energies were produced at the tail of the explosion 
interval, this fact could not explain such a big difference in the time spread due to the sharp 
fall off of events. The spread of the arrival times in the IMB detector, which cannot observe 
neutrino energy lower than 20 MeV, is nearly 6 seconds, nearly half of the value observed at 
Kamiokande II, but still much longer than the expected duration of the supernova 
explosion. Furthermore, in spite of its lower statistics,  the Baksan data, collected with a 
threshold energy about halfway between the IMB and the Kamiokande II ones, give a time 
spread about halfway between the values obtained by the last two laboratories. 

In the next sections, it will be shown that also this smooth change of the time 
spreads with the detectors' thresholds is an effect related to the non-zero masses of the 
neutrinos arriving on the Earth.

\section{Time delay of massive neutrino} 
	Let us look now at the time delay in the arrival time of a non-zero mass neutrino in 
comparison to that of a massless one. If the mass is exactly zero, the time of flight for 
arriving on the Earth from the Supernova is the same for all the neutrinos. It is 
     $$ T_0 = L_{SN} / c = 1.7 \cdot 10^5 \quad years $$

\noindent where  $L_{SN}$  is the distance of the Supernova from the Earth, and c  is the light speed in 
vacuum. However if the mass m is not zero, then the time of flight is  
 
$$ T_m = \frac{L_{SN}}{c \cdot \sqrt{1-(m/E)^2}} \sim \frac{L_{SN}}{c} \cdot \{1+\ 1/2 \cdot (m/E)^2 \} $$
 
	The difference of these two values, {\it i.e.} the delay in the arrival of a  neutrino with 
mass m  in comparison to a massless one, is 
 
$$  \Delta T_m = T_m -T_0 \approx  1/2 \cdot T_0 \cdot m^2/E^2 $$

Numerically, a neutrino of energy 5 MeV should delay about one second if the mass is 3 eV 
and about 10 seconds if the mass is 10 eV.
 
\section{ Data plotted in the diagram: \mbox{\boldmath $\Delta T$}
  vs. \mbox{\boldmath $1/E^2$}} 
	The best way to see if such a mass effect exists at all consists in plotting $\Delta T$, the 
arrival time delay of each event, {\it versus} $1/E^2$, the reciprocal of the observed squared energy.  
Before doing that, it is noted that the supernova explosion takes a finite (non zero) time. 
Thus the delay in the arrival time of the nth observed event reads  

 $$\Delta T_n = \tau_n + T_0m_\alpha^2 / 2E_n^2  ,$$

\noindent where index $\alpha$ specifies the kind of the arrived neutrino and $\tau_n$   denotes the time when the 
neutrino was produced since the start of the supernova explosion.  It should be observed 
that $\Delta T_n$  increases with n and that it depends on the observed energy 
$E_n$ as well as on the 
mass - for the moment unknown - of the {\it n} th observed neutrino. Assuming that the 
distribution of the emission time $\tau_n$
 is narrow, from the previous equation it follows that, in a 
diagram $\Delta T$ {\it versus} $1/E^2$,  the data points ($\Delta T_n , 1/E_n^2$) should clearly form a straight line 
with a positive and finite slope,  if the neutrinos have a non-zero mass. In fact, the slope is 
simply proportional to the squared mass value of the neutrino. Moreover, if there are more 
than one mass state, in the diagram one should observe as many straight lines as the number 
of the different masses not equal to zero. If the supernova explosion takes a finite time, then 
the points ($\Delta T_n, 1/E_n^2$), relevant to neutrinos with mass $m_\alpha$, must lie inside a stripe 
delimited by two parallel straight lines with slope equal to $T_0 m_\alpha^2/2$. Noting that the $\tau_n$ 
distribution is the same for each mass group, because the explosion is the same for each 
mass state, and the intersection of each stripe with the time axis will also be the same. 

         In Fig.3 we report all the data observed by the three laboratories mentioned above. 
This procedure is quite legitimate. In fact, the different location of the laboratories on the 
Earth is at most responsible for a delay of a tenth of second, while the uncertainty in the 
time zero among the different detectors can be estimated in less than 2 tenths of second 
from the time separation in the arrival times of the first events at each detector. (Of course, 
these two uncertainties disappear if we confine ourselves to consider the data collected by a 
single laboratory.)  It should be also noted that Fig.3 was obtained by setting the origin of 
the time axis at the arrival time delay of the first observed event. As a consequence of this 
choice, assuming the explosion instantaneous, each straight line will intersect the time axis 
slightly before the origin. Fig.3 clearly shows two well separated grouping of the observed 
events. Moreover, each group obeys the conditions required by the massive nature of 
neutrinos, namely:

1)	a less energetic event arrives later than a more energetic one within experimental 
errors,

2)	the events are well distributed inside a narrow strip close to a straight line,

3)	the slope of this line is positive and is proportional to the squared mass of the 
corresponding neutrino,

4)	the straight line intersects the time axis just before the arriving time delay of the first 
event,

5)	the linear course grained distributions of the points inside each strip are similar for 
the two strips, and the intersections of the strip with the time axis are roughly less  
than 1 s. 

The linear fits of the two grouping of the only Kamiokande II data, characterized by a lower 
energy threshold and a lower experimental error, yield the following two mass values \cite{10}
 
        $$      m_1  =  3.4     \pm     0.6     eV$$
    and
        $$      m_2  =   22. \pm 4.      eV$$
	
\noindent The data plot in this diagram two separated linear mass groups are
very clearly visible. The order statistics applied to the relation between order
of arriving time and that of its energy of the events, for the first and the
second mass group separately, gives us both more than 90\% confidence.
However this statistics tests only monotone relation between two orders without
discriminating two opposite sense of relations; in our case, physically good
sense or nonsense (lower energy event arrives later or earlier). Adding this
effect the above obtained value becomes as more than 95\%.
Furthermore taking into account the other physically necessary conditions; 1)
not only monotonous relation in good sense but also the linear form in this
diagram, further 2) the fact that this line should hit a little bit before the
arriving time of the first event, and also 3) the consistency with all the other
data obtained by independent apparatus, the confidence level should be very high.

Very rough time distribution of the neutrino emission at the Supernova can be
easily obtained and shown in Fig.4 both $m_1$ and $m_2$ separately. The
distribution for $m_2$ comes out very much wider. But 
two events far from the center of the explosion corresponding to a little before
the time zero, one at about the beginning and the other twards the ending, have
the time errors of several seconds due to the large observation error of low
energy and also due to the large slope of the mass-fit line. Both events are the
last arriving in the detector of KAM II. Taking into account this
effect, the both emission time distribution are consistent as the same.

\section{Discussion}
             The previous plot and the mass values obtained by its analysis deserve some 
remarks:

1)  It should be noted the usefulness of plotting the delay in the arrival times of neutrinos
 {\it versus} the reciprocal squared energy. It is this plot that clearly shows the separation of the 
observed events into two groups. The precise determination of the arrival times and the 
$10 \sim 30$\% errors on the energies of the observed neutrinos allows us to fit the separate groups of 
events with two straight half lines which intersect the vertical axis near the origin. In this 
respect, the assumption that neutrinos are massive particles plays the fundamental role. On 
the one hand, it explains the separate groupings of the events. On the other hand, it also 
explains the rather large spread in the arrival times. Moreover, the energy distributions in the 
two groups of events look similar (see Fig.5 in Ref.\cite{10}). 

2) For each group of events, consider 
now the deviation, along the time axis, of each event from the best-fitted straight line. The 
distributions of these deviations, that correspond to $\tau_n$ of the preceeding section, are also 
quite similar between the two groups and narrower than one second. The similarity of the 
distributions of energy and time deviation for the two groups of events is consistent with the 
fact that the neutrino of each event, produced during the supernova explosion with a well 
defined flavor state, generates  well separated mass states after its long journey towards the 
Earth. Last but not least, the best fit of the two almost linear groupings allows us to 
determine the masses of at least two neutrinos (for the third neutrino mass, see the 
discussion in the next section) and not simply their mass difference.
 
3)  It is important to combine in a single plot all the data collected by the three laboratories. 
It has already been explained the reason why this is legitimate.  We stress now the usefulness 
of this procedure. The data collected by the IMB apparatus refer to rather energetic 
neutrinos owing to the high energy threshold of the detector. Thus, no wonder that the IMB 
data show only one grouping of these events so as to make the observation of two masses  
impossible. The data collected by the Baksam apparatus refer to neutrinos of lower energy. 
But, due to the small size of this detector, the total number (only 5) of the detected events is 
too small to make any mass effect clearly visible. However, this effect becomes clearly visible 
in the plot including Kamiokande II data. In fact, the Kamiokande II detector performed 
much better thanks to its larger size and to the enormous efforts made to purify the water of 
the apparatus so as to ensure a very good transparency to faint light (the Kamioka 
underground mine has a wonderful water spring that produces excellent pure water in 
enormous quantity, but this is only a small part of the excellent function of this detector) and 
it has a much lower energy threshold. In this way, the low-energy events detected by the 
Kamiokande II apparatus make the separation of the events into two groups quite evident.
 
4)  The previous findings no longer require to assume that the duration of the supernova 
explosion is longer than predicted in order to explain the wide spread observed in the arrival 
times, and that a large fraction of low energy neutrinos is produced toward the end of the 
supernova's explosion. Furthermore, an explosion of 10 s or more would make it difficult to 
explain the reason why the observed events lie within two different strips. On the contrary, 
the small time spread of the two bunches of straight lines in the plot indicates that the 
explosion lasted less than 1 s. A shorter explosion also agrees with the observation that only 
7 events, instead of the expected 50 ones, have energy higher than 10 MeV. 

5) The mechanism that could show particle mass in the method is extremely simple
and fundamental rule in physics: the relation between velocity and energy of a
particle that depends on only one variable, its mass. It is clear that 
the method
does not depend on the initial condition of the particle at leaving from the
source, but only on the observation of the particle on the Earth. Also
often physical value and character of a particle appears in several different 
independent modes. In this case each result of all the different mechanism 
should be physically important.

\section{The third mass} 
	Neutrinos are expected to have three mass states as well as three flavors. Then, one 
rightly wonders why only two mass values are clearly visible in the reported plot.  We are 
therefore obliged to speculate on the possible reason hiding the appearance of a third mass 
value. 

1)  If the third mass is zero and it interacts in the electronic flavor channel, it cannot escape 
from observation. In this case, the associated events should concentrate in a narrow group 
around the time of the first arrived event. This group cannot experimentally get lost. Thus 
one concludes that the existence of a massless neutrino with electronic flavor is impossible.

2)  If we assume that the third mass is very large, e.g. around 100 eV or bigger, the arrival 
times of these neutrinos would spread over a time range wide several minutes or more. 
Then, it would be quite awkward to distinguish the real supernova signals from the 
background events. However, if the mass is so large, it should easily be observed through 
decay processes of charged leptons or in neutrino oscillation phenomena. In conclusion, the 
existence of a quite large mass neutrino appears unlikely. 

3)  The missing mass state could have a mixing angle of nearly 90 with the electronic flavor 
state ({\it i.e.} a very small component of electronic flavor state).  Thus such neutrinos could 
hardly produce a visible electronic event. Moreover, the highest energy value observed for 
the neutrinos, arrived from the supernova, is 40 MeV, and energies higher than this value are 
quite unlikely according to theoretical estimates of the explosion power \cite{1}. Thus, this kind 
of neutrinos would hardly have enough energy to produce other leptons as a muon 
(105.66 MeV) or a much heavier tauon (1784.2 MeV) via other flavor channels. Therefore, 
in this case,  any  value for  the third mass is possible, including the zero value. Besides, 
if such a case were true, the nearly equal number of events in the two mass groups indicates 
that these two mass states have roughly the same mixing angles with the electronic flavor 
state and therefore the maximum mixing would be favored. 

4)  The mass value of the third neutrino is fairly close to one of the two observed values 
(within a few of eV).  Thus, either 

 $$ m_3 \approx m_1  $$
or
 $$ m_3 \approx m_2      $$

The second possibility is favored by the recent results on the atmospheric neutrino 
oscillations, observed at Super-Kamiokande \cite{11,12} by comparing the neutrino behaviors 
over distances of the Earth diameter, since the observed squared mass difference turned out 
to be about $10^{-3}(eV)^2$.
 
5)  It is also mentioned that some theoretical models on neutrino mixing predict only two 
observable mass states \cite{13}. 

\section{Conclusion} 
	Neutrinos emitted from a very distant supernova, once arrived on Earth, could reveal 
all their physical features, as masses and mixing angles among both mass- and flavor-states, if 
all the flavors can be observed by the detectors and if all the three mass states are well 
separated.

	In the case of Supernova LMC-'87A, however, it was impossible to get this full 
information partly because the only electronic flavor could be observed due to the low 
energy of neutrinos and partly because two masses only are visibly separated. Even though 
no information about the mixing angles among mass and flavor states can be obtained, this 
explosion was an extremely important opportunity to prove the existence of at least two 
massive neutrino kinds and to get a reliable estimate of their masses.  This was made possible 
by the plot of the time delay in the arrival times  {\it versus} the reciprocal squared energy of the 
events. The plot also showed the full consistency of the data observed by the three 
laboratories. For the third neutrino mass, one can only say either that its value is close to one 
of the observed ones or that it is not visible because the neutrino has a very small electron 
flavor. 

\section*{Acknowledgments}
	The author would like to thank Prof. Salvino Ciccariello, Prof. Ferruccio Ferruglio 
and Dr. Marco Laveder for discussions, comments and interesting suggestions on this work.

 \clearpage
 \begin{figure}
 \caption{
   Expected Time Distribution of Neutrinos
    produced at LMC'87-A}
     \label{TimeDistr} 
\epsfig{file=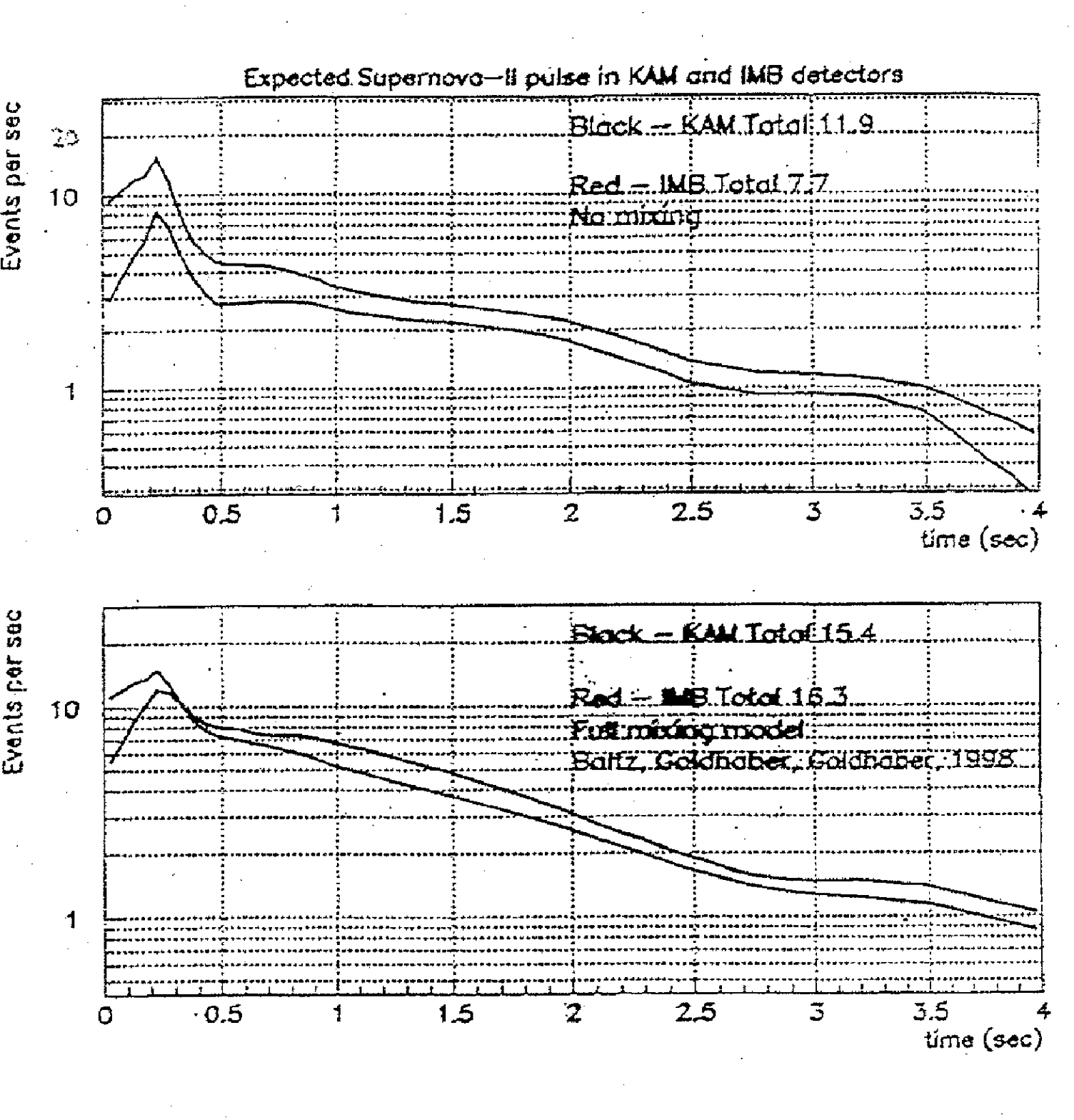,width=0.94\textwidth} 
\end{figure}
\newpage 


\begin{figure} 
\vspace{2cm}
\caption{ Relation between the lowest energy of events
     and the spread of arriving time of events}
\label{Relation}
\mbox{\hspace*{-2cm}\epsfig{file=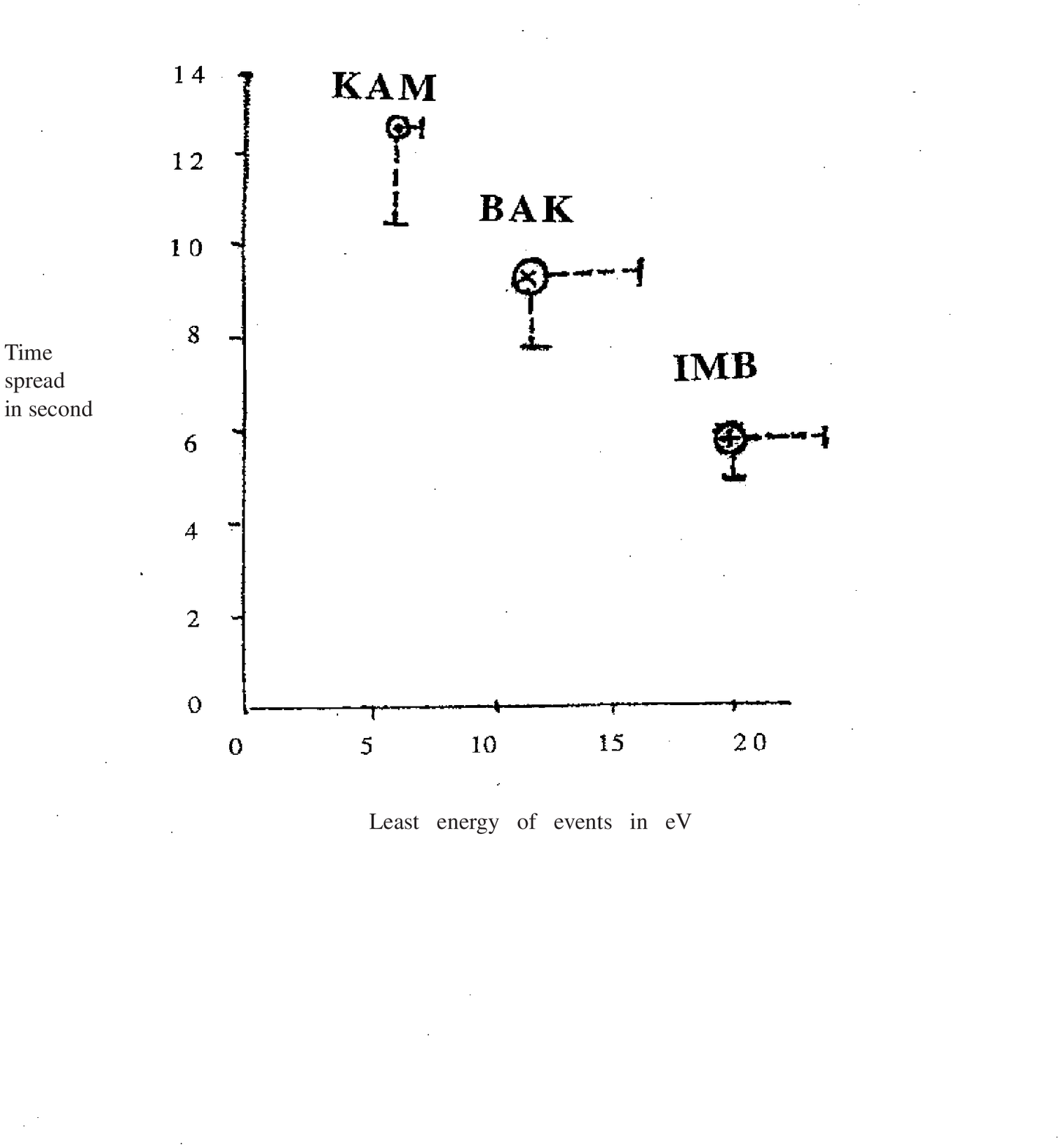,
               width=50cm, height=20cm,
               bbllx=49bp,bblly=510bp,bburx=507bp,bbury=700bp,clip=}} \\
\end{figure}

\newpage
\begin{figure}
\caption{
     T vs. $1/E^2$ Plot of the observed events}
\label{Plot}
\epsfig{file=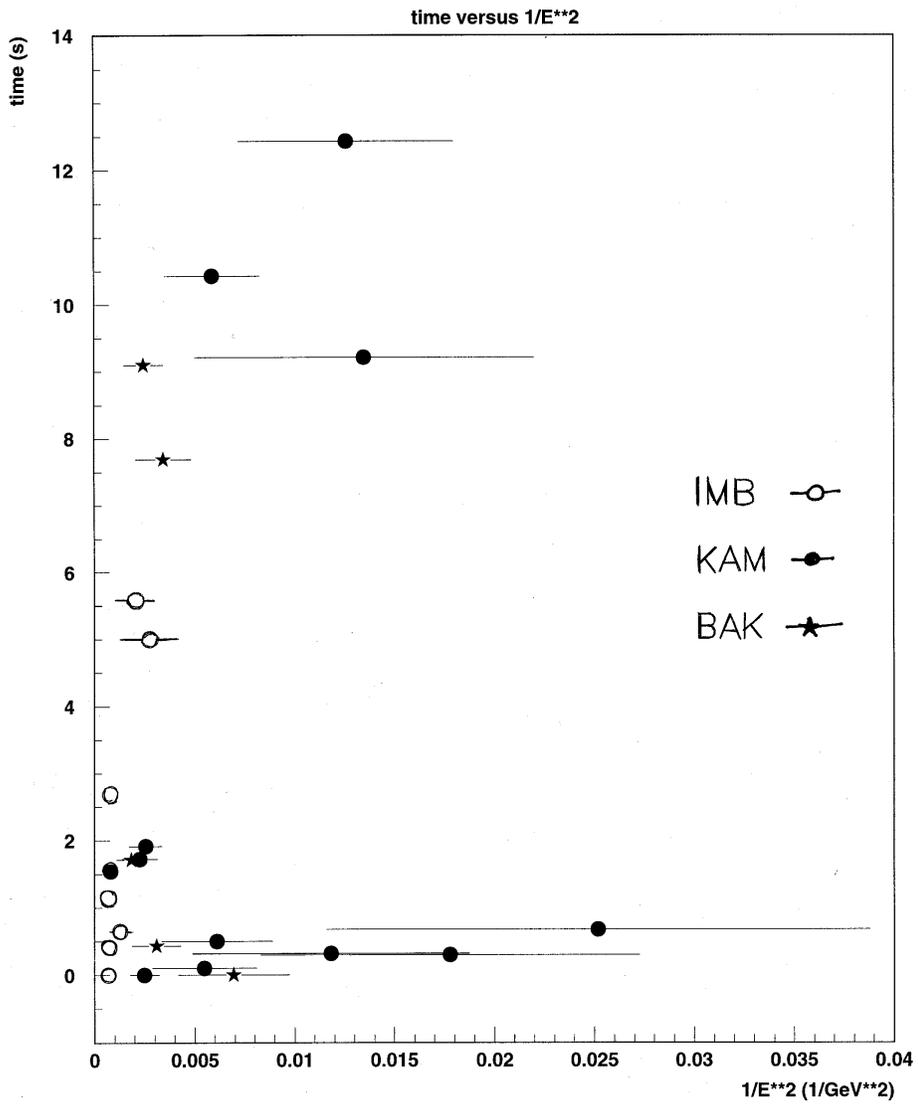,width=0.94\textwidth}
\end{figure}
\newpage
 \begin{figure}
 \caption{Emission time distribution of neutrinos at the Supernova}
     \label{Distribution} 
\mbox{\hspace*{-2.0cm}\epsfig{file=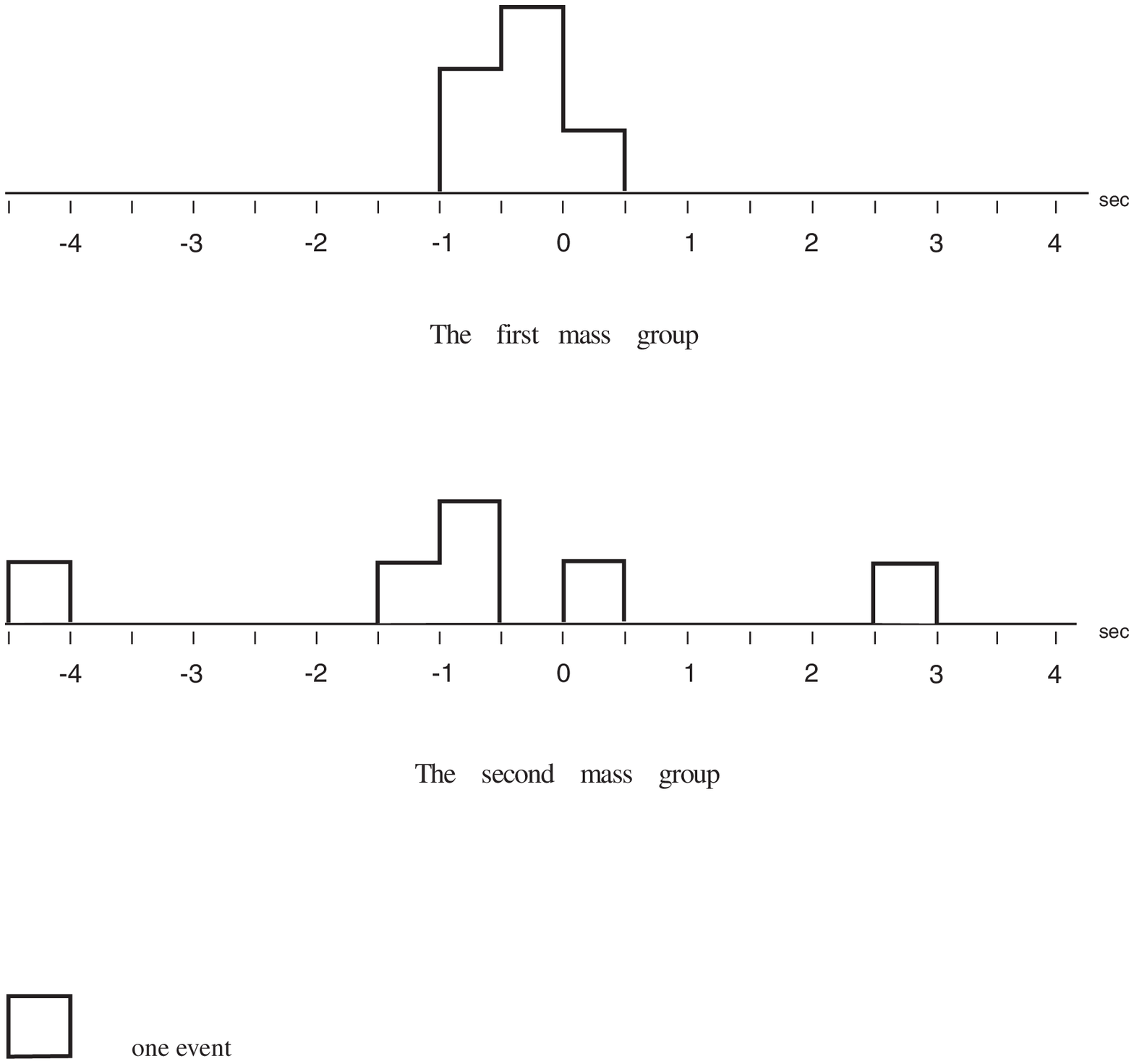,width=20truecm}}
\end{figure}

\begin{thebibliography}{99}
 
  \bibitem{1}H. Hirata et al., Phys. Rev. Letters B, Vol. 58 ('87), 1490. 
  \bibitem{2}H. Hirata et al., Phys. Rev. D Vol. 38 ('88), 448. 
  \bibitem{3}R. M. Bionta et al., Phys. Rev. Letters Vol. 58 (87), 1494. 
  \bibitem{4}C. B. Bratton et al., Phys. Rev. D Vol. 37 ('88), 3351. 
  \bibitem{5}E. N. Alexeyev et al., Phys. Rev. Letters B, Vol. 205 ('88), 209.
  \bibitem{6}J. N. Bahcall,  A. Dar and T. Piran, Nature Vol. 326 ('87), 133.
  \bibitem{7}J. R. Wilson "Numerical Astrophysics"(1986), Ed.s Centrella J. et al.,
		 (Jones \& Bartlett, Boston).
  \bibitem{8}J. R. Wilson et al. N.Y. Acad. Sci. Vol. 470 ('86), 267.
  \bibitem{9} M. Goldhaber and M. Divanat, Neutrino Telescope.  Venice ('99). 
\bibitem{10}H. Huzita,  Modern Physics Letters A, Vol. 2 ('87), 905. 
\bibitem{11}Y. Fukuda et al., Phys. Rev. Letters, Vol. 82 ('99), 2644. 
\bibitem{12}Y. Fukuda et al., Phys. Letters B, Vol. 467 ('99), 185. 
\bibitem{13}S. M. Bilenky, C. Giunti and W. Grimus, Prog. Part. Nucl. Phys. Vol 43 ('99), 1 
\end{thebibliography}
\end{document}